\documentclass[a4paper]{PoS}
\DeclareMathAlphabet{\mathcal}{OMS}{cmsy}{m}{n}
\usepackage{amsmath}
\DeclareMathAlphabet\mathrsfso{U}{rsfso}{m}{n}

\title{Studying nucleon structure via Double Deeply Virtual
Compton Scattering (DDVCS)}

\ShortTitle{Double Deeply Virtual
Compton Scattering (DDVCS)}

\author{\speaker{Shengying Zhao}\thanks{Supported by the China Scholarship Council (CSC) and the French Centre National de la Recherche Scientifique (CNRS).} \\
        Institut de Physique Nucl\'{e}aire d'Orsay, CNRS-IN2P3, Universit\'{e} Paris-Sud \& Paris-Saclay, 91406 Orsay, France\\
        E-mail: \email{zhao@ipno.in2p3.fr}}

\abstract{Study of the structure and dynamics of the nucleon has been deeply renewed with the advent of a parameterization of the partonic structure of the nucleon in terms of the Generalized Parton Distributions (GPDs). Encoding the correlations between the elementary constituents of the nucleon, GPDs allow a 3-dimensional imaging of the nucleon from the dynamical link between the transverse position and the longitudinal momentum of partons. 
Double Deeply Virtual Compton Scattering (DDVCS) corresponds to the scattering from the nucleon of a virtual photon that finally generates a lepton pair $eN \rightarrow eN \gamma^* \rightarrow eN l\bar{l}$ where the final leptons can be either an $e^+e^-$ or a $\mu^+\mu^-$ pair. The virtuality of the final photon allows to investigate in a decorrelated way the initial and transferred momentum dependences of the GPDs. This unique feature of DDVCS is of relevance, among others, for the determination of the transverse parton densities and the distribution of nuclear forces.
This proceeding discusses preliminary model-predicted DDVCS experimental projections at JLab12 and indicates the impact of potential DDVCS experiments.
}

\FullConference{23rd International Spin Physics Symposium - SPIN2018 -\\
		10-14 September, 2018\\
		Ferrara, Italy}

\begin{document}

\section{Introduction}

There are essentially three experimental golden channels for direct measurements of GPDs: the electroproduction of a photon $eN\rightarrow eN\gamma$ which is sensitive to the deeply virtual Compton scattering (DVCS) amplitude, the photoproduction of a lepton pair $\gamma N\rightarrow l\bar{l}N$ which is sensitive to the timelike Compton scattering (TCS) amplitude, and the electroproduction of a lepton pair $eN\rightarrow eNl\bar{l}$ which is sensitive to the double deeply virtual Compton scattering (DDVCS) amplitude. Only the latter provides the framework necessary for an uncorrelated measurement of a GPD ($\xi',\xi,t$) as a function of both scaling variable $\xi'$ and $\xi$ \cite{ref1,ref2}. The former two reactions cannot entirely serve the purpose of testing the angular momentum sum rule \cite{ref3} due to the reality of the final- or initial-state photons, which leads to the restriction $\xi'=\pm \xi$. For instance, the Compton form factors (CFF) $\mathcal{H}$ associated with the GPD $H$ and accessible in DVCS cross section or beam spin asymmetry experiments can be written
\begin{eqnarray}
\relax\mathcal{H}(\xi'=\xi,\xi,t)=\sum_{q}e_q^2&\bigg\{&\mathcal{P}\int_{-1}^1dx~H^q(x,\xi,t)\bigg[\frac{1}{x-\xi}+\frac{1}{x+\xi}\bigg]
\nonumber\\
&&-i\pi\big[H^q(\xi,\xi,t)-H^q(-\xi,\xi,t)\big]\bigg\}
\label{eq1}
\end{eqnarray}
where the sum runs over all parton flavors with elementary electrical charge $e_q$, and $\mathcal{P}$ indicates the Cauchy principal value of the integral. While the imaginary part of the CFF accesses the GPD values at $\xi'=\pm \xi$, it is clear from Eq.~\ref{eq1} that the real part of the CFF is a more complex quantity involving the convolution of parton propagators and the GPD values out of the diagonals $\xi'=\pm \xi$, that is a domain that cannot be resolved unambiguously with DVCS experiments. Because of the virtuality of the final state photon, DDVCS provides a way to circumvent the DVCS limitation, allowing to vary independently $\xi'$ and $\xi$. Considering the same GPD $H$, the corresponding CFF for DDVCS process writes
\begin{eqnarray}
\mathcal{H}(\xi',\xi,t)=\sum_{q}e_q^2&\bigg\{&\mathcal{P}\int_{-1}^1dx~H^q(x,\xi,t)\bigg[\frac{1}{x-\xi'}+\frac{1}{x+\xi'}\bigg]
\nonumber\\
&&-i\pi\big[H^q(\xi',\xi,t)-H^q(-\xi',\xi,t)\big]\bigg\}
\label{eq2}
\end{eqnarray}
providing access to the scaling variable $\xi' \neq \xi$.

The DDVCS process is most challenging from the experimental point of view due to the small magnitude of the cross section and requires high luminosity and full exclusivity of the final state. Moreover, the difficult theoretical interpretation of electron-induced lepton pair production when detecting the $e^+ e^-$ pairs from the decay of the final virtual photon, hampers any reliable experimental study. Taking advantage of the energy upgrade of the CEBAF accelerator, it is proposed to investigate the electroproduction of $\mu^+ \mu^-$ di-muon pairs and measure the beam spin asymmetry of the exclusive $ep\rightarrow e'p'\gamma^* \rightarrow e'p'\mu^+\mu^-$ reaction in the hard scattering regime \cite{intent,intent2,ref6}.

\begin{figure}[t]
\centering
\includegraphics[width=.4\textwidth]{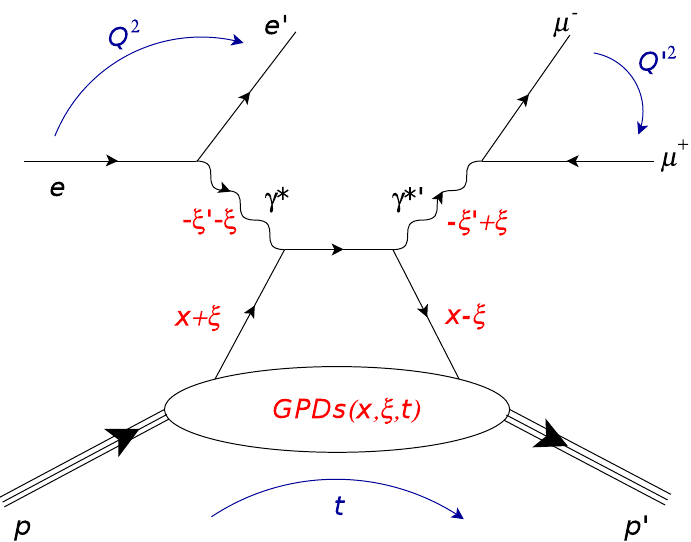} 
\caption{The handbag diagram symbolizing the DDVCS direct term with di-muon final states.}
\label{fig1}
\end{figure}

At sufficiently high virtuality of the initial space-like virtual photon and small enough four-momentum transfer to the nucleon with respect to the photon virtuality ($-t \ll Q^2$), DDVCS can be seen as the absorption of a space-like virtual photon by a parton of the nucleon, followed by the quasi-instantaneous emission of a time-like virtual photon by the same parton, which finally decays into a di-muon pair (Fig.~\ref{fig1}). $Q^2$ and $Q'^2$ represent the virtuality of the incoming space-like and outgoing time-like photons. The scaling variable $\xi'$ and $\xi$ write
\begin{eqnarray}
\xi' = \frac{Q^2-Q'^2+t/2}{2Q^2/x_\text{B}-Q^2-Q'^2+t}~~\text{and}~~
\xi  = \frac{Q^2+Q'^2}{2Q^2/x_\text{B}-Q^2-Q'^2+t}
\label{xipxi}
\end{eqnarray}
from which one obtains
$\xi' = \xi  \frac{Q^2-Q'^2+t/2}{Q^2+Q'^2} .$
This relation indicates that $\xi'$, and consequently the CFF imaginary part, is changing sign about $Q^2=Q'^2$, which procures a strong testing ground of the universality of the GPD formalism. 

In this proceeding, the feasibility of a DDVCS experiment at JLab 12 GeV is discussed. Section \ref{sec2} describes DDVCS kinematics and experimental observables. Section \ref{sec3} reports model-predicted experimental projection at a certain luminosity with ideal detectors. Preliminary conclusions about this study are drawn in the last section.

\section{Kinematics and experimental observables}
\label{sec2}

\begin{figure}[t]
\centering
\includegraphics[width=.49\textwidth]{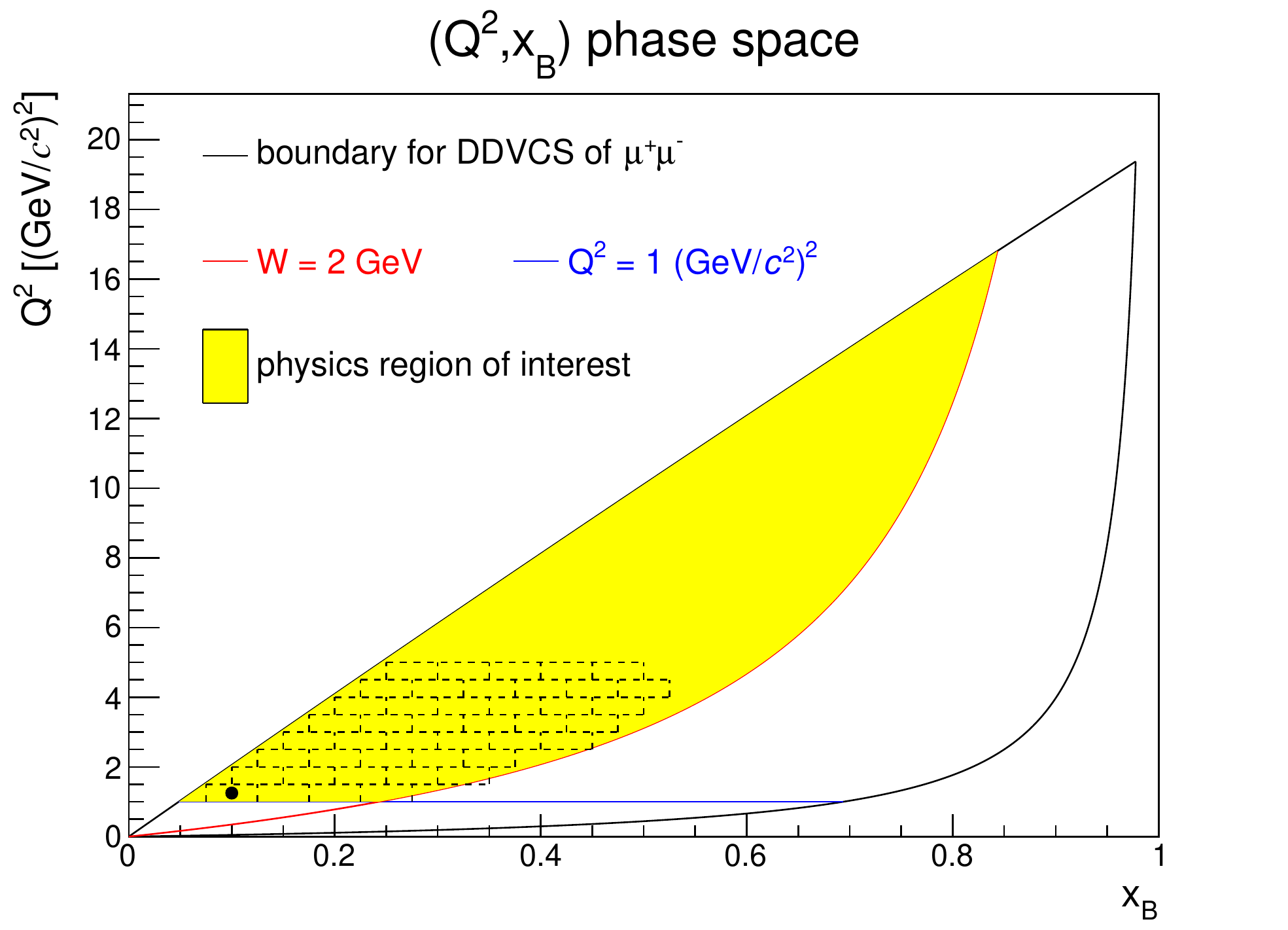}
\includegraphics[width=.49\textwidth]{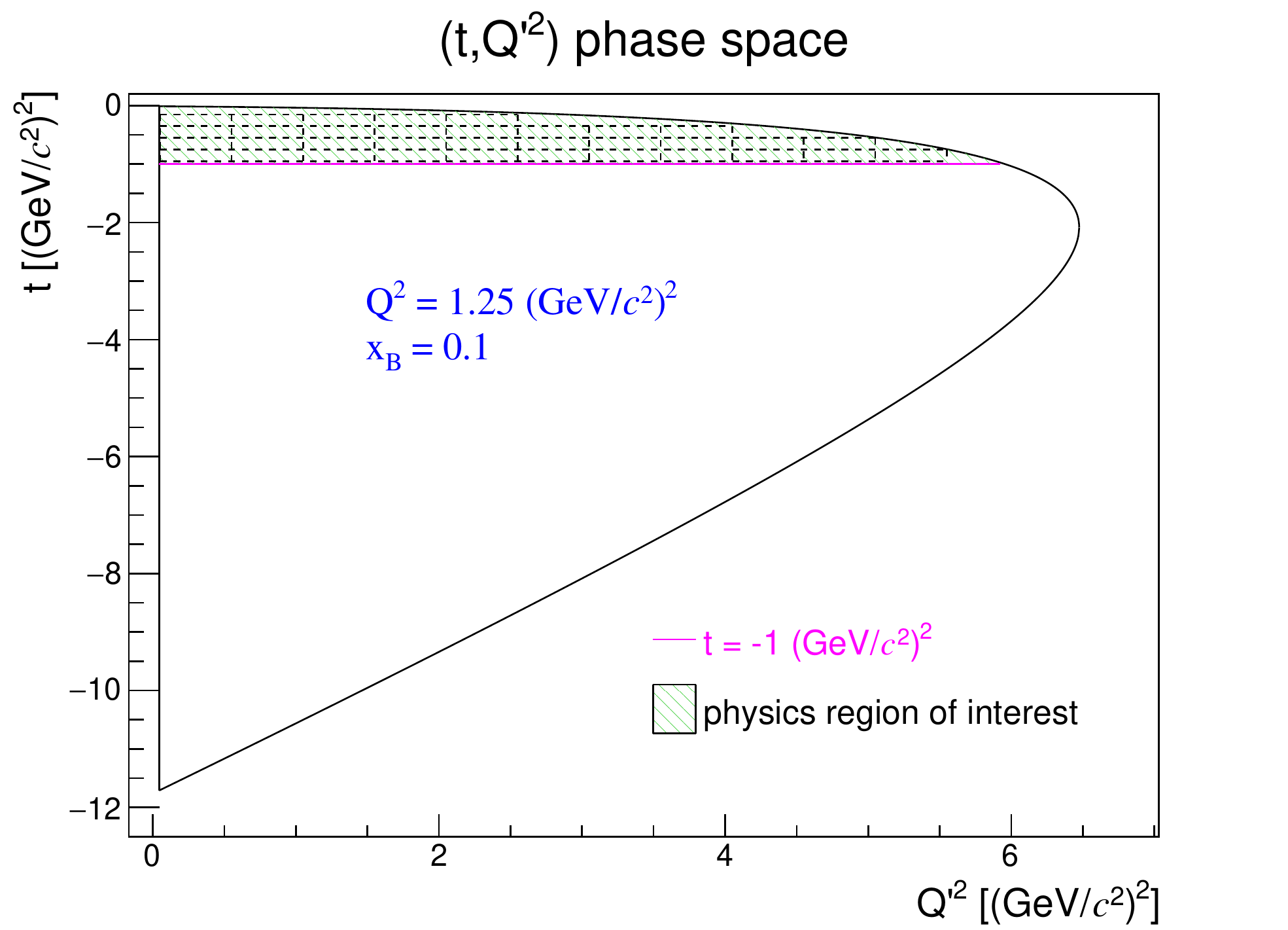} 
\caption{Kinematics phase spaces: left panel shows the $(Q^2,~x_\text{B})$ phase space with physics inspired cuts. The shaded area represents the physics region of interest, and the point represents $(Q^2=1.25~($GeV$/c^2)^2,~x_\text{B}=0.1)$ whose correlated $(t,~Q'^{2})$ phase space is shown in the right panel together with the region of interest (shaded area).}
\label{fig2}
\end{figure}

The following kinematics cuts to ensure applicability of
the GPD formalism have been applied: the center-of-mass energy $W>2$ GeV to ensure the deep inelastic scattering regime; $Q^2>1~($GeV$/c^2)^2$ to ensure the reaction at parton level; $t >-1~($GeV$/c^2)^2$ to support the factorization regime; and $Q'^2>(2m_\mu)^2$ to ensure the production of a di-muon pair. Fig.~\ref{fig2} shows the DDVCS allowed $(Q^2,~x_\text{B})$ phase space with kinematics cuts and one $(t,~Q'^{2})$ phase space at a specific $(Q^2,~x_\text{B})$ set. In the allowed region represented by the shaded area, kinematics bins with uniform widths have been chosen. $\Delta Q^2=0.5~($GeV$/c^2)^2$, $\Delta x_\text{B}=0.05$, $\Delta t=0.2~($GeV$/c^2)^2$ and $\Delta Q'^2=0.5~($GeV$/c^2)^2$, which is sufficiently small to allow a first-order estimation of the integral cross section. As a preliminary study of DDVCS, only the bins at $Q^2<5~($GeV$/c^2)^2$ have been studied, where the cross section is supposedly larger than at high $Q^2$. As a consequence, 664 four-dimensional bins have been considered. The bins boundary are shown in Fig.~\ref{fig2} as the dashed lines.

\begin{figure}[b]
\centering
\includegraphics[height=.2\textwidth]{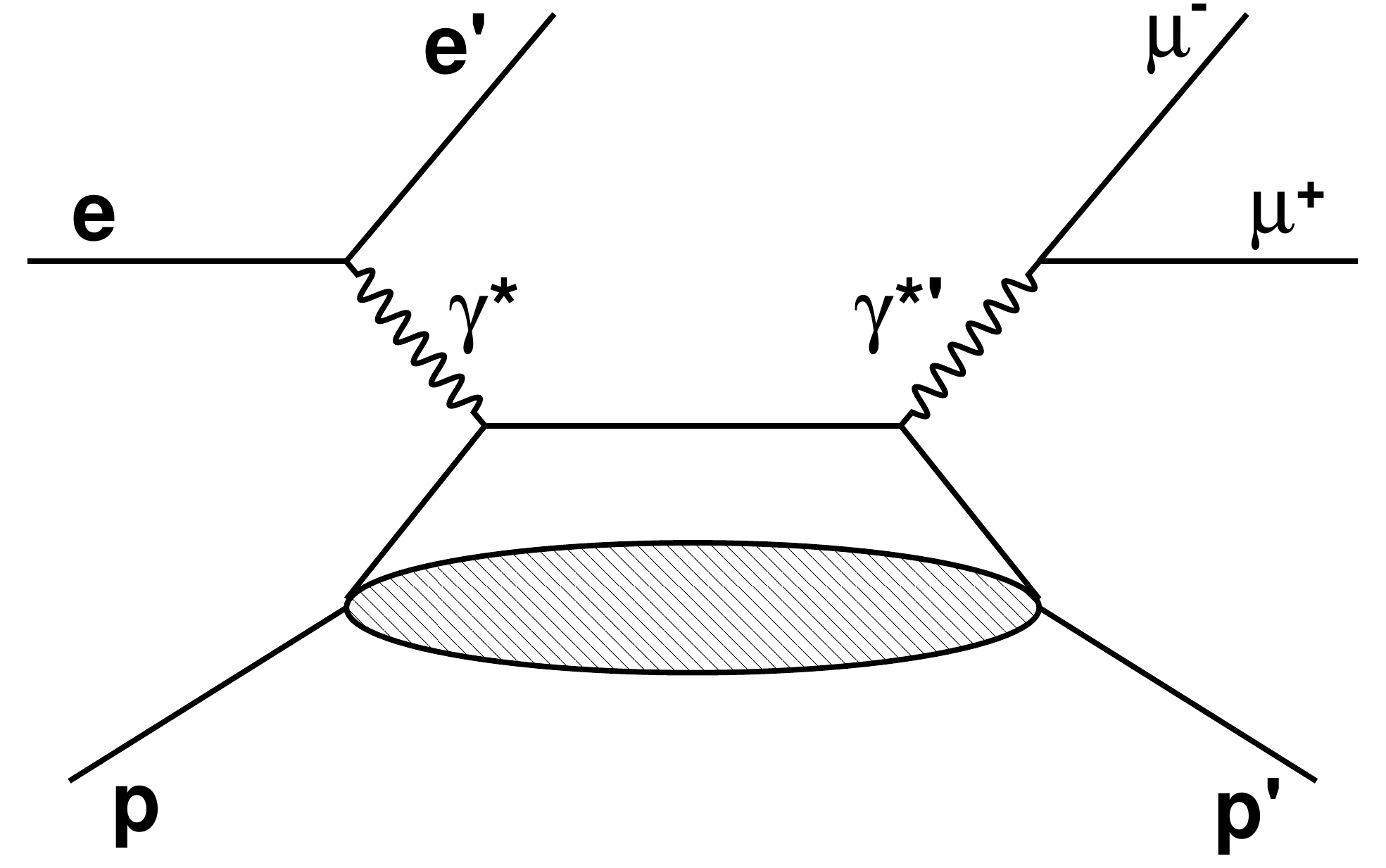} ~~~~~~~~
\includegraphics[height=.2\textwidth]{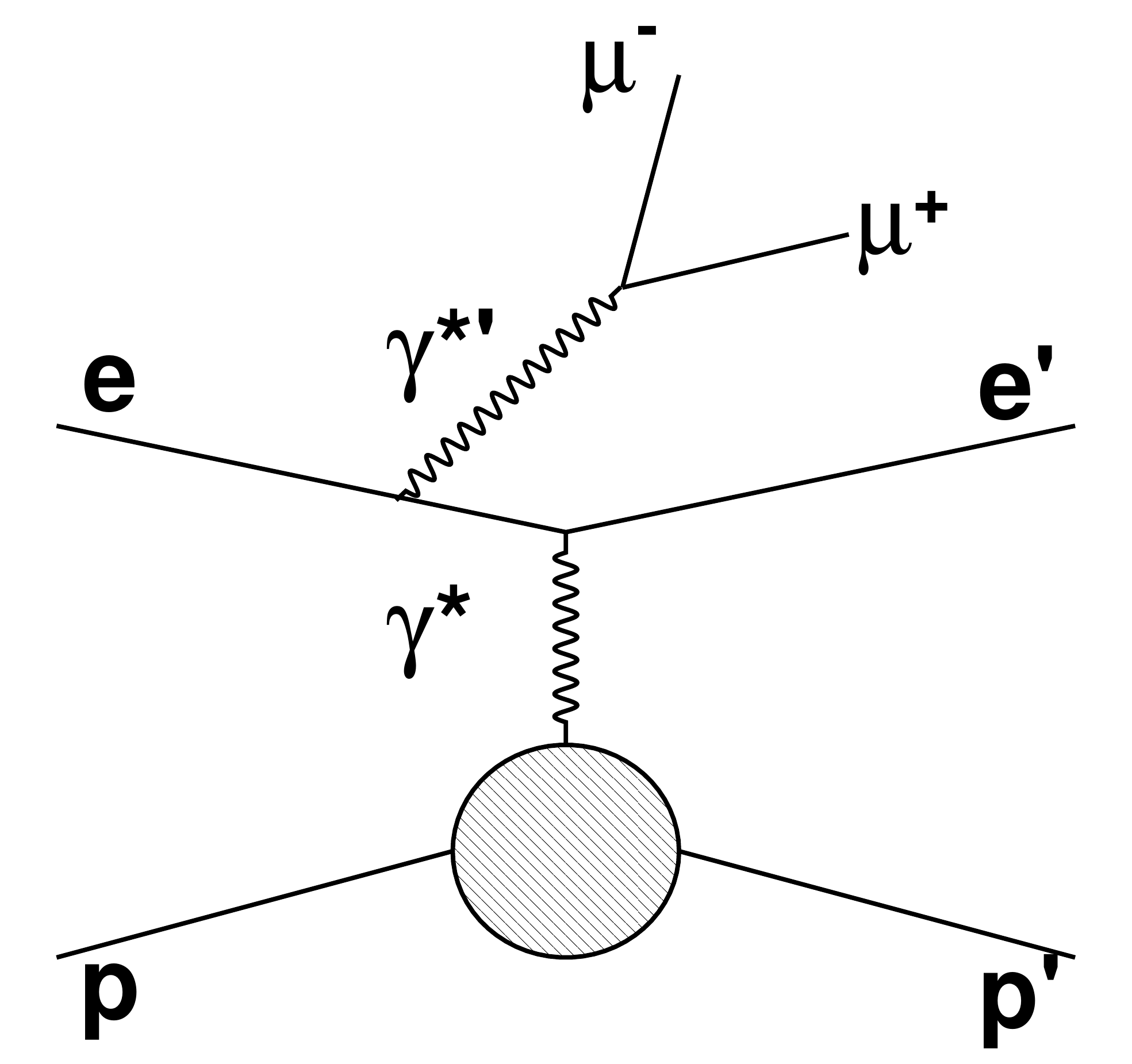} ~~~~~~~~
\includegraphics[height=.2\textwidth]{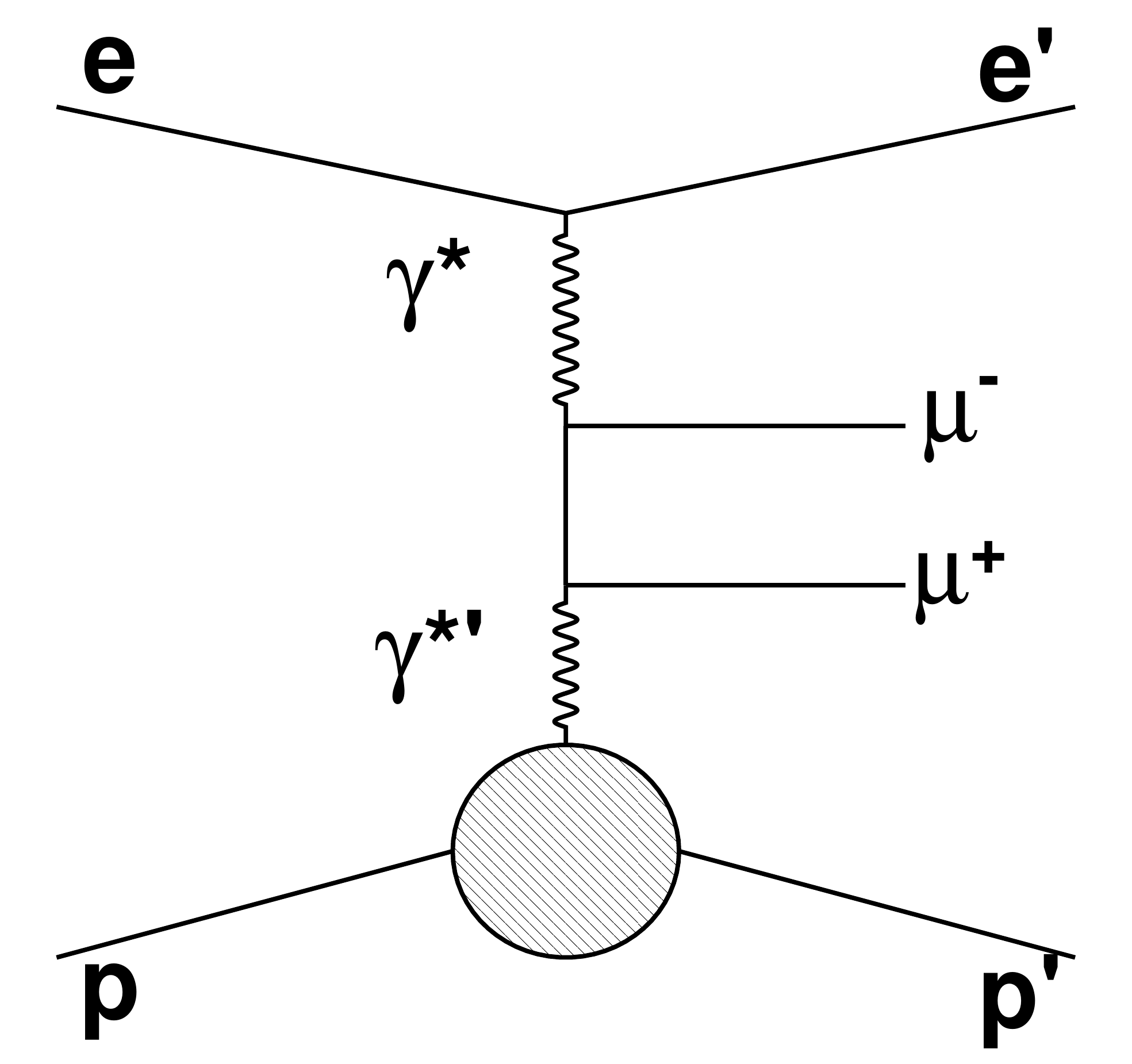} 
\caption{Subprocesses contributing to electroproduction of a di-muon pair including DDVCS (left) and two kinds of Bethe-Heitler processes, i.e. BH$_1$ (middle) and BH$_2$ (right).}
\label{fig3}
\end{figure}

The lepton-pair electroproduction process consists of three interfering elementary mechanisms, depicted in Fig. \ref{fig3}, with implied crossed contributions. The 7-fold differential cross section is proportional to the square of the total amplitude that is the coherent sum of the three processes, i.e. $d^7\sigma/(dQ^2dx_BdtdQ'^2d\phi d\Omega_\mu)$ $\propto |\mathcal{T}_\text{DDVCS}+\mathcal{T}_{\text{BH}_1}+\mathcal{T}_{\text{BH}_2}|^2$. We consider in this study the 5-fold cross section, integrating over the muon solid angle. The integration leads to the vanishing of the interference contributions originated from the BH$_2$ amplitude: $d^5\sigma/(dQ^2dx_BdtdQ'^2d\phi) \propto |\mathcal{T}_\text{DDVCS}+\mathcal{T}_{\text{BH}_1}|^2+|\mathcal{T}_{\text{BH}_2}|^2$ \cite{ref2,ref7}. Though partial information is sacrificed, this simplification offers an easier understanding of this totally unexplored reaction. The cross section without target polarization can be described in terms of different contributions
\begin{equation}
\sigma_{P}^{e}=\sigma_{\text{BH}_1}+\sigma_{\text{BH}_2}
+\sigma_\text{DDVCS}+P\widetilde{\sigma}_\text{DDVCS}+(-e)\left( \sigma_{\text{INT}_1}+P\widetilde{\sigma}_{\text{INT}_1} \right)
\label{eq3}
\end{equation}
where, to simplify the notation, $\sigma$ stands for the 5-fold differential cross section, $e$ is the lepton beam electric charge, and $P$ is the polarization of the beam. The subscript INT$_1$ represents the interference terms between the DDVCS and the BH$_1$ amplitudes. BH terms are calculable since the nucleon form factors are well-known at small $t$. DDVCS terms are bi-linear in CFFs, while interference terms are linear. $\sigma_\text{DDVCS}$ and $\sigma_{\text{INT}_1}$ are sensitive to the real part of CFFs, while $\widetilde{\sigma}_\text{DDVCS}$ and $\widetilde{\sigma}_{\text{INT}_1}$ are sensitive to the imaginary part.

Considering polarized positron and electron beams, single contributions can be separated from the three experimental observables: unpolarized cross section with electron beam ($\sigma_\text{UU}$), beam spin cross section difference with polarized electron and positron beam ($\Delta\sigma_\text{LU}$), and beam charge cross section difference ($\Delta\sigma^\text{C}$). From Eq. \ref{eq3},
\begin{equation}
\left\{
\begin{aligned}
&\sigma_\text{UU}=\frac{1}{2}\left(\sigma_{+}^-+\sigma_{-}^-\right)
&&=\sigma_{\text{BH}_1}+\sigma_{\text{BH}_2}
+\sigma_{\text{DDVCS}}
+\sigma_{\text{INT}_1},
\\
&\Delta\sigma_\text{LU}=\frac{1}{4}\left[\left(\sigma_{+}^--\sigma_{-}^-\right)-\left(\sigma_{+}^+-\sigma_{-}^+\right)\right]
&&=\widetilde{\sigma}_{\text{INT}_1},
\\
&\Delta\sigma^\text{C}=\frac{1}{4}\left[\left(\sigma_{+}^-+\sigma_{-}^-\right)-\left(\sigma_{+}^++\sigma_{-}^+\right)\right]
&&=\sigma_{\text{INT}_1}.
\end{aligned}
\right.
\label{eq4}
\end{equation}
It is difficult to extract CFFs from DDVCS term of bi-linear combination, $\Delta\sigma^\text{C}$ therefore provides pure interference term. The imaginary part of CFFs can be extracted from $\Delta\sigma_\text{LU}$, which provides directly the information for GPDs. In addition, we can also obtain pure $\sigma_\text{DDVCS}$ when combining $\sigma_\text{UU}$ and $\Delta\sigma^\text{C}$, and $\widetilde\sigma_\text{DDVCS}$ when combining $\Delta\sigma_\text{LU}$ and the electron beam spin cross section difference. The experimental projections of these observables have been performed, and is discussed in the next section.

\section{Projections}
\label{sec3}
The projections have been performed in the ideal situation that all the particles of the final state can be detected with 100\% efficiency. The count-rate calculation was done for a luminosity $\mathrsfso{L}=10^{37}~\text{cm}^{-2}\text{s}^{-1}$ considering 100 days running time equally distributed between each lepton beam charge. The number of events, for each five-dimensional bin ($Q^2,~x_\text{B},~t,~Q'^{2}~\text{and}~\phi$), was determined following
\begin{eqnarray}
N=\frac{d^5\sigma}{dQ^2dx_BdtdQ'^2d\phi} \cdot\Delta Q^2 \cdot\Delta x_B \cdot\Delta t \cdot\Delta Q'^2 \cdot\Delta\phi \cdot\mathrsfso{L} \cdot T,
\label{eq5}
\end{eqnarray}
where the differential cross section has been calculated with the VGG model \cite{refVGG} set at the central values of each four-dimensional bin at the beam energy of 11 GeV. Besides, 24 bins in $\phi$ 15$^\circ$-wide have been considered.

\begin{figure}
\centering
\includegraphics[width=1\textwidth]{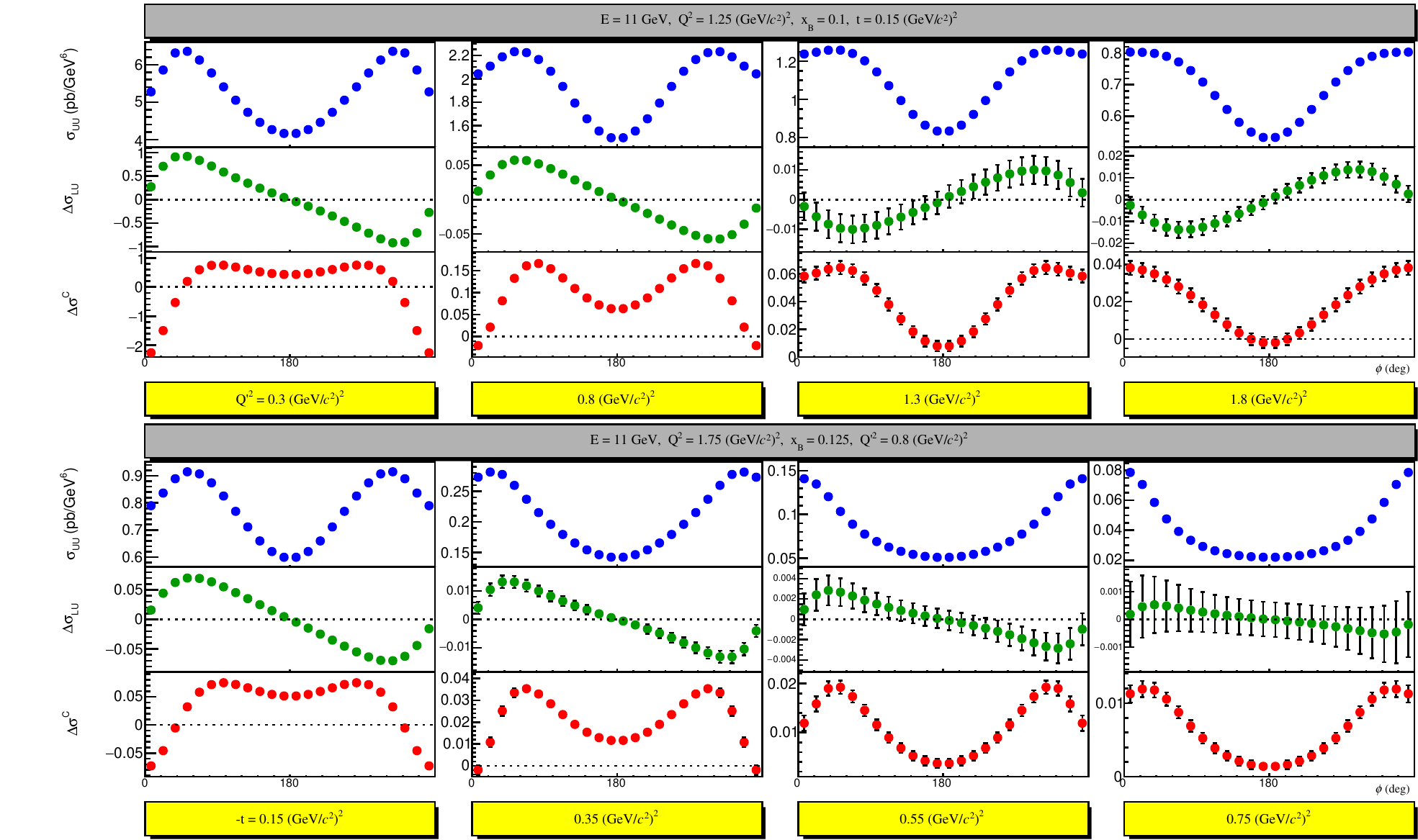}
\caption{The upper half of the figure shows the VGG projections at $Q^2 =1.25~(\text{GeV}/c^2)^2$, $x_\text{B}=0.1$, $-t=0.15~(\text{GeV}/c^2)^2$ and  $Q'^2=0.3,~0.8,~1.3~\text{and}~1.8~(\text{GeV}/c^2)^2$ (left to right). The panels of the top row show the unpolarized cross section, the middle panels show the beam-spin cross section difference, and the bottom panels show the beam-charge cross section difference. Note that each panel has its own y-axis scale. The bottom half shows the ones at $Q^2 =1.75~(\text{GeV}/c^2)^2$, $x_\text{B}=0.125$, $Q'^2=0.8~(\text{GeV}/c^2)^2$ and  $-t=0.15,~0.35,~0.55~\text{and}~0.75~(\text{GeV}/c^2)^2$ (left to right).}
\label{fig4}
\end{figure}

Fig.~\ref{fig4} shows the observables with statistic errors at some different $Q'^2$ and a set of fixed ($Q^2,~x_\text{B}~\text{and}~t$) as a function of $\phi$ (upper half). The cross section decreases generally as $Q'^2$ increases, since the process at $Q'^2=0$ is equivalent to the DVCS process having one less electromagnetic vertex. $\Delta\sigma_\text{LU}$ in the $Q^2>Q'^2$ and $Q^2<Q'^2$ regions has opposite signs due to the antisymmetric property of GPD \cite{ref7}. The bottom half shows the ones at some different $t$ and a set of fixed ($Q^2,~x_\text{B}~\text{and}~Q'^2$). The cross section and precision decrease as $-t$ increases.

\begin{figure}
\centering
\includegraphics[width=.32\textwidth]{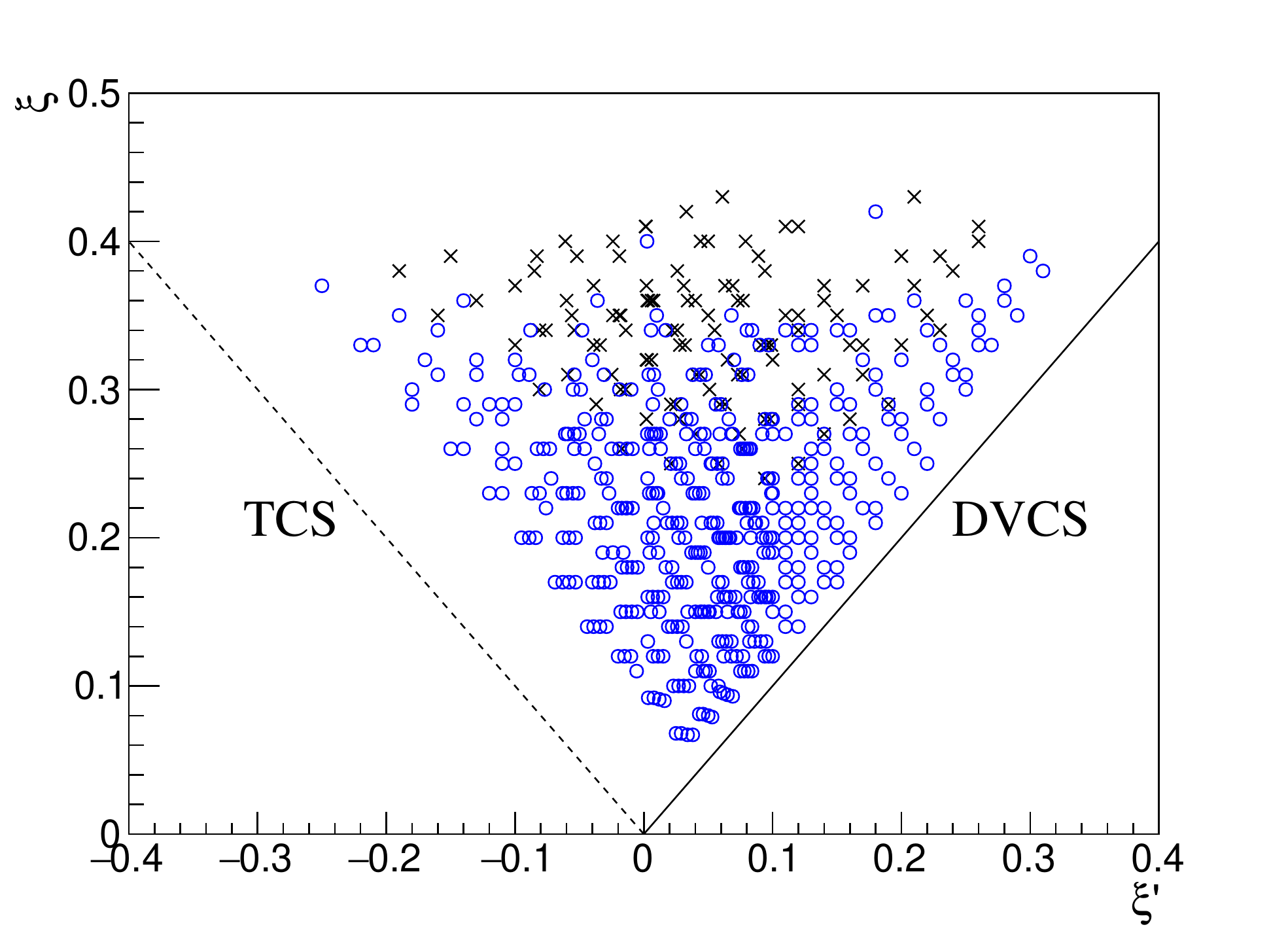} 
\includegraphics[width=.32\textwidth]{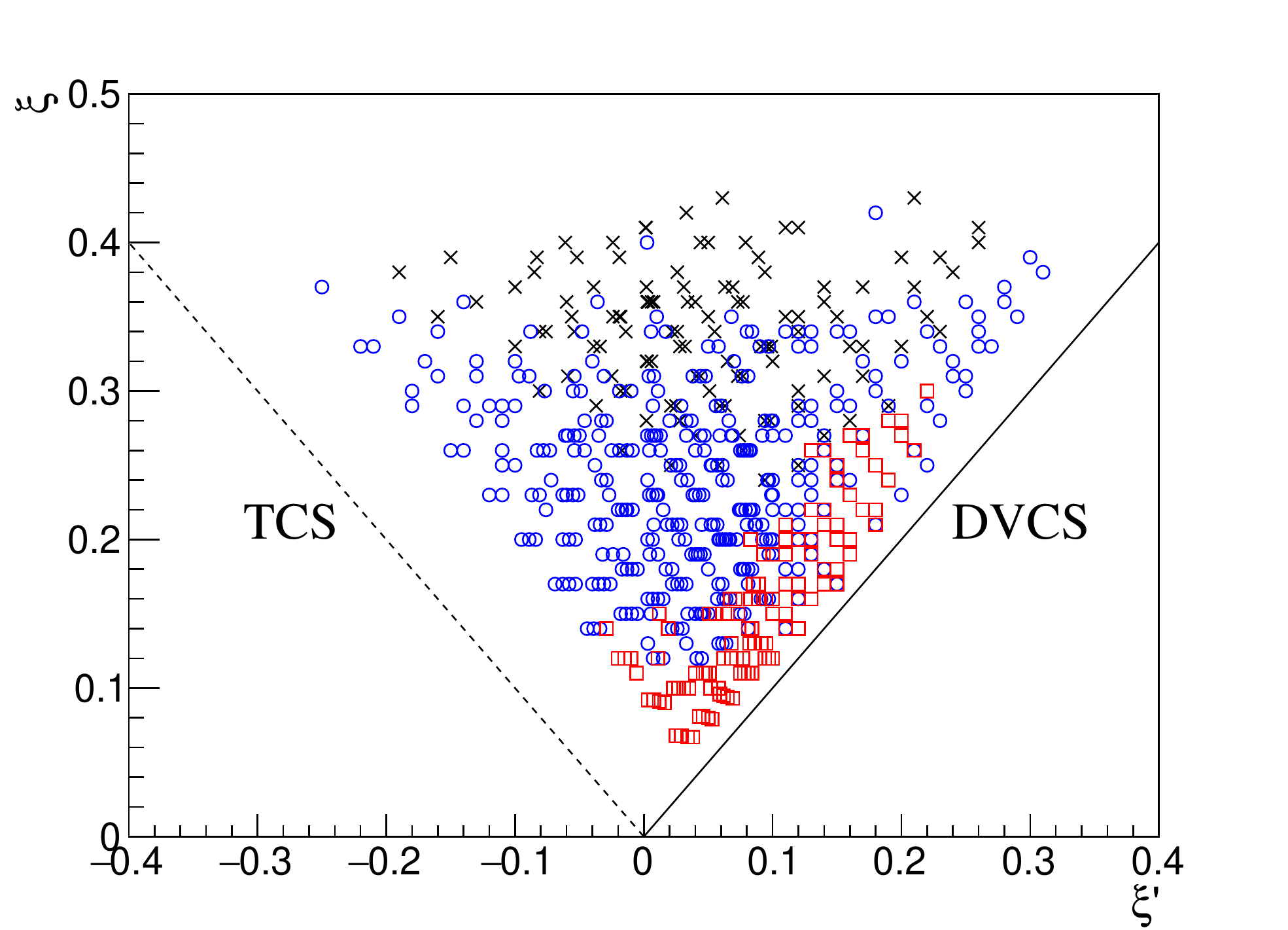} 
\includegraphics[width=.32\textwidth]{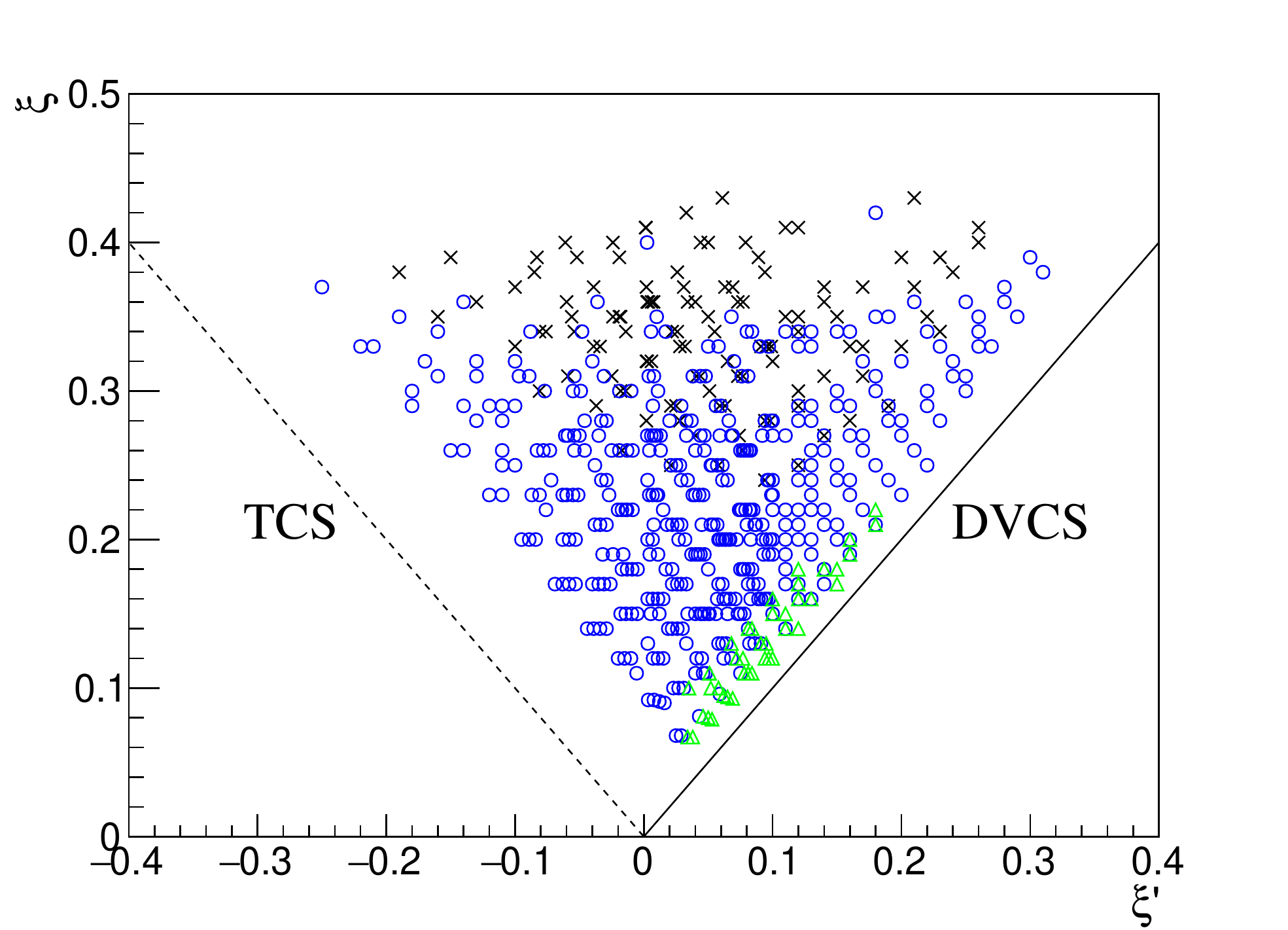} 
\caption{CFFs phase space: the solid line indicates $\xi'=\xi$ or $Q'^2=0$ that is the DVCS correlation, the dashed line indicates $\xi'=-\xi$ or $Q^2=0$ that is the TCS correlation, and the colored markers represent the successful four-dimensional bins of DDVCS process (Eq.~1.2). The blue open circles indicate the bins where $\sigma_\text{UU}$ has the relative error less than 10\%.  Some of them, represented by the red open squares, have $\Delta\sigma^\text{C}$ with the relative error less than 10\%, and a few of them, represented by green open triangles, have the $\Delta\sigma_\text{LU}$ at the same level of accuracy. The black crosses indicate the failed bins where all the three observables have the relative errors greater than 10\%.}
\label{fig5}
\end{figure}

Fig.~\ref{fig5} shows the correlated location of all the four-dimensional bins in the CFFs phase space $(\xi',~\xi)$. Among the 664 bins, 82\% have $\sigma_\text{UU}$ with a relative error less than 10\% , 22\% have  $\Delta\sigma^\text{C}$ of the same quality, and only 7\% for $\Delta\sigma_\text{LU}$ in the vicinity of the DVCS diagonal.

\section{Conclusion}
\label{sec4}

The model-predicted projections of a DDVCS experiment indicate a high degree of feasibility at a challenging luminosity with exclusive final states completely detected. The unpolarized cross section with very small statistics error can be obtained. Although the beam charge cross section difference has less precision, a better extraction of the real part of CFFs can be performed. The beam spin cross section difference can be obtained accurately only at a few specific kinematics, but it is the most powerful tools to directly access the totally unexplored GPDs phase space, otherwise inaccessible. An additional feature supporting the importance of this observable is the sign change of the beam spin cross section difference as $Q'^2$ becomes larger than $Q^2$. This behaviour is a strong prediction of the GPD formalism \cite{ref7}, and consequently provides a stringent test for experimental investigations.

Due to the strong sensitivity to $Q'^2$ and $t$ as well as the small cross section and insufficient statistics at high values, the binning strategy will be adapted in the next phase of DDVCS exploration covering the whole kinematic phase space, and finally extracting the CFFs.

\section*{Acknowledgement}
I would like to express my appreciation to my thesis supervisor, E. Voutier, for the guidance of this work, also to M. Guidal, S. Niccolai, and M. Vanderheaghen for helpful discussions.

\end{document}